\documentclass[11pt,twoside]{article}
\usepackage{asp2010}

\resetcounters

\begin{document}

\title{Discovery of a stripped red giant core in a bright eclipsing binary
star }
\author{P.~F.~L.~Maxted$^{1}$, D.~R.~Anderson$^{1}$, M.~R.~Burleigh$^2$, 
A.~Collier Cameron$^3$, U.~Heber$^{4}$, B.~T.~G\"{a}nsicke$^5$, S.~Geier$^4$,
T.~Kupfer$^{4}$, T.~R.~Marsh$^{5}$, G.~Nelemans$^6$, S.~J.~O'Toole$^7$,
R.~H.~{\O}stensen$^8$, B.~Smalley$^{1}$, R.~G. West$^{9}$, S.~Bloemen$^{8}$
\affil{$^1$Astrophysics Group,  Keele University, Keele, 
      Staffordshire ST5 5BG}
\affil{$^2$Department of Physics and Astronomy, University of Leicester, University
Road, Leicester LE1 7RH}
\affil{$^3$SUPA, School of Physics and Astronomy, University of St.\
Andrews, North Haugh,  Fife, KY16 9SS, UK}
\affil{$^4$Dr. Karl Remeis-Observatory \& ECAP, Astronomical Institute,
Friedrich-Alexander University Erlangen-Nuremberg,} Sternwartstr.~7,
D~96049 Bamberg, Germany}
\affil{$^5$Department of Physics, University of Warwick, Coventry, CV4 7AL }
\affil{$^6$Department of Astrophysics, IMAPP, Radboud University Nijmegen, PO Box
9010, 6500 GL Nijmegen, The Netherlands }
\affil{$^7$Australian Astronomical Observatory, PO Box 296, Epping, NSW, 1710,
Australia}
\affil{$^8$Institute of Astronomy, K.U.Leuven, Celestijnenlaan 200D, 3001, Heverlee,
Belgium}
\affil{$^9$Department of Physics and Astronomy, University of Leicester,
University Road, Leicester LE1 7RH}

\begin{abstract}
 We report the serendipitous discovery from WASP archive photometry of a
binary star in which an apparently normal A-type star (J0247$-$25\,A) eclipses
a smaller, hotter subdwarf star (J0247$-$25\,B). The kinematics of
J0247$-$25\,A show that it is a blue-straggler member of the Galactic
thick-disk.  We present follow-up photometry and spectroscopy from which we
derive approximate values for the mass, radius and luminosity for
J0247$-$25\,B  assuming that J0247$-$25\,A has the mass appropriate for a
normal thick-disk star. We find that the properties of J0247$-$25\,B are well
matched by models for a red giant stripped of its outer layers and currently
in a shell hydrogen-burning stage. In this scenario, J0247$-$25\,B  will go on
to become a low mass white dwarf ($M\sim0.25M_{\odot}$) composed mostly of
helium. J0247$-$25\,B can be studied in much greater detail than the handful
of pre helium white dwarfs (pre-He-WD) identified to-date. These results have
been published by \citet{2011arXiv1107.4986M}. We also present a preliminary
analysis of more recent observations of J0247$-$25 with the UVES spectrograph,
from which we derive much improved masses for both stars in the binary. We
find that both stars are more massive than expected and that J0247$-$25\,A
rotates sub-synchronously by a factor of about 2. We also present lightcurves
for 5 new eclipsing pre-He-WD subsequently identified from the WASP archive
photometry, 4 of which have mass estimates for the subdwarf companion based on
a pair of radial velocity measurements. 

\end{abstract}

\section{Discovery of J0247$-$25}

The WASP survey (Wide Angle Search for Planets, \citealt{2006PASP..118.1407P})
uses two instruments to monitor the brightness of millions of stars in both
hemispheres. Each instrument has 8 e2V CCD cameras with 200\,mm f/1.8 Canon
lenses to produce images covering approximately $8^{\circ}\times8^{\circ}$ on
the sky per camera. Two 30s exposures are obtained on selected fields 
 every 5\,--\,10 minutes every clear night. The strategy is optimised for the
detection of planetary transits for stars with V$\approx$ 9-13. The techniques
used to identify planetary transits in the WASP data are also very effective
at identifying  eclipsing binary stars.  One star flagged as an eclipsing
binary star as part of this process was 1SWASP~J024743.37$-$251549.2
(J0247$-$25 hereafter). The WASP lightcurve of this star is shown as a
function of orbital phase with the period 0.6678\,d in Fig.\ref{lc}. The
shape and depths of the eclipses in this lightcurve show that the feature at
phase 0 is the total eclipse of a smaller but hotter star by its larger and
cooler companion. The catalogue photometry available show that
the larger star (J0247$-$25\,A),  which contributes $\sim$90\% of the optical
light, is an A-type star, so we obtained follow-up observations to determine
the nature of the smaller, hotter star. Photometry with the SAAO 1.0-m
telescope (Fig.\ref{lc}) confirmed our interpretation of the WASP lightcurve.
Spectroscopy with a variety of instruments was used to confirm the mid-A
spectral type of J0247$-$25\,A and to measure its the spectroscopic orbit
(Fig.~\ref{spec}).  We used the lightcurve model
 EBOP \citep{1981psbs.conf..111E,1981AJ.....86..102P} to analyse the WASP
and SAAO 1.0-m lightcurves. The surface brightness ratio we derive from the
lightcurve models can be combined with the observed V and K$_{\rm S}$
magnitudes of J0247$-$25 to estimate the effective temperatures 
T$_{\rm eff,A}\approx 8060$K and T$_{\rm eff,B}\approx 13400$K.
The surface gravity of J0247$-$25\,B, $\log g_B = 4.76\pm0.05$,
 can be derived directly from the parameters of the lightcurve model and the
mass function. In Fig.~\ref{SafferTlogg} we compare these values of 
T$_{\rm eff,B}$ and $\log g_B$ to the effective temperatures and surface
gravities of 298 faint blue stars observed by \cite{1997ApJ...491..172S}. It
is clear that J0247$-$25\,B is unusually cool given its surface gravity and
sits well below the main sequence (long-dashed lines) and the zero-age
horizontal branch (short-dashed lines). 

\articlefigure{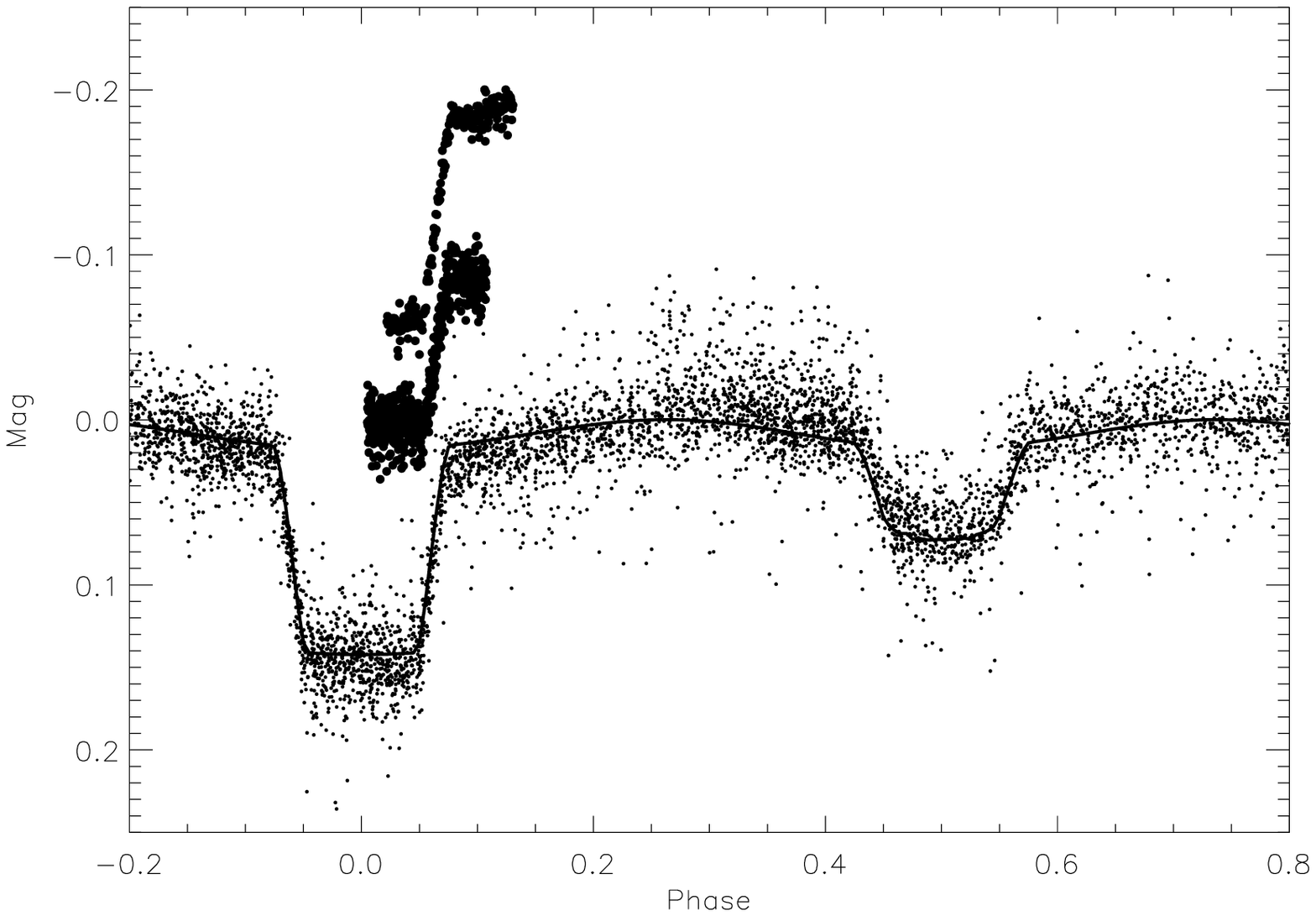}{lc}{Lightcurves of J0247$-$25. From
bottom-to-top: WASP white-light photometry with lightcurve model fit,
SAAO 1.0-m I$_{\rm C}$-band and V-band.}

\articlefiguretwo{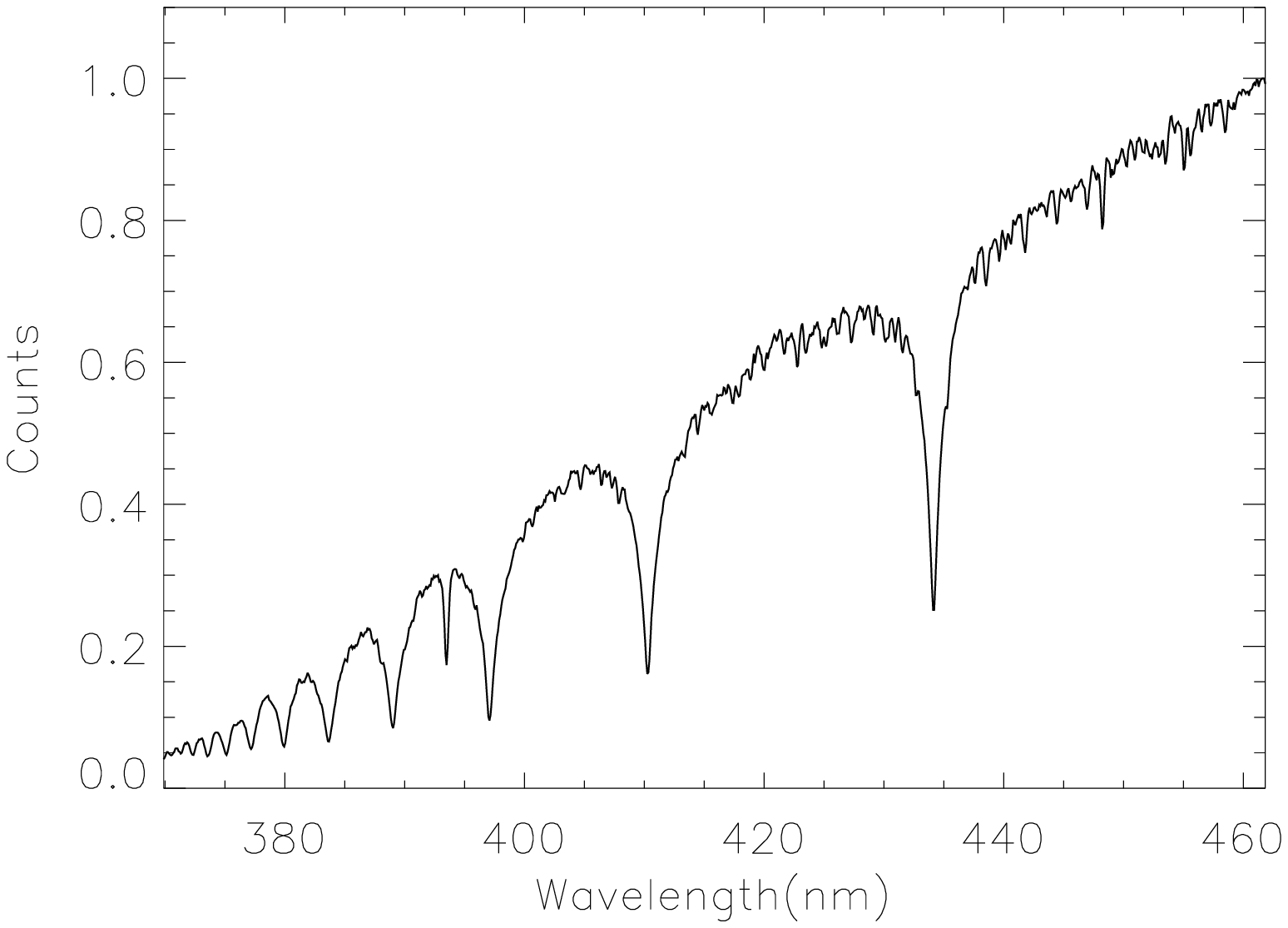}{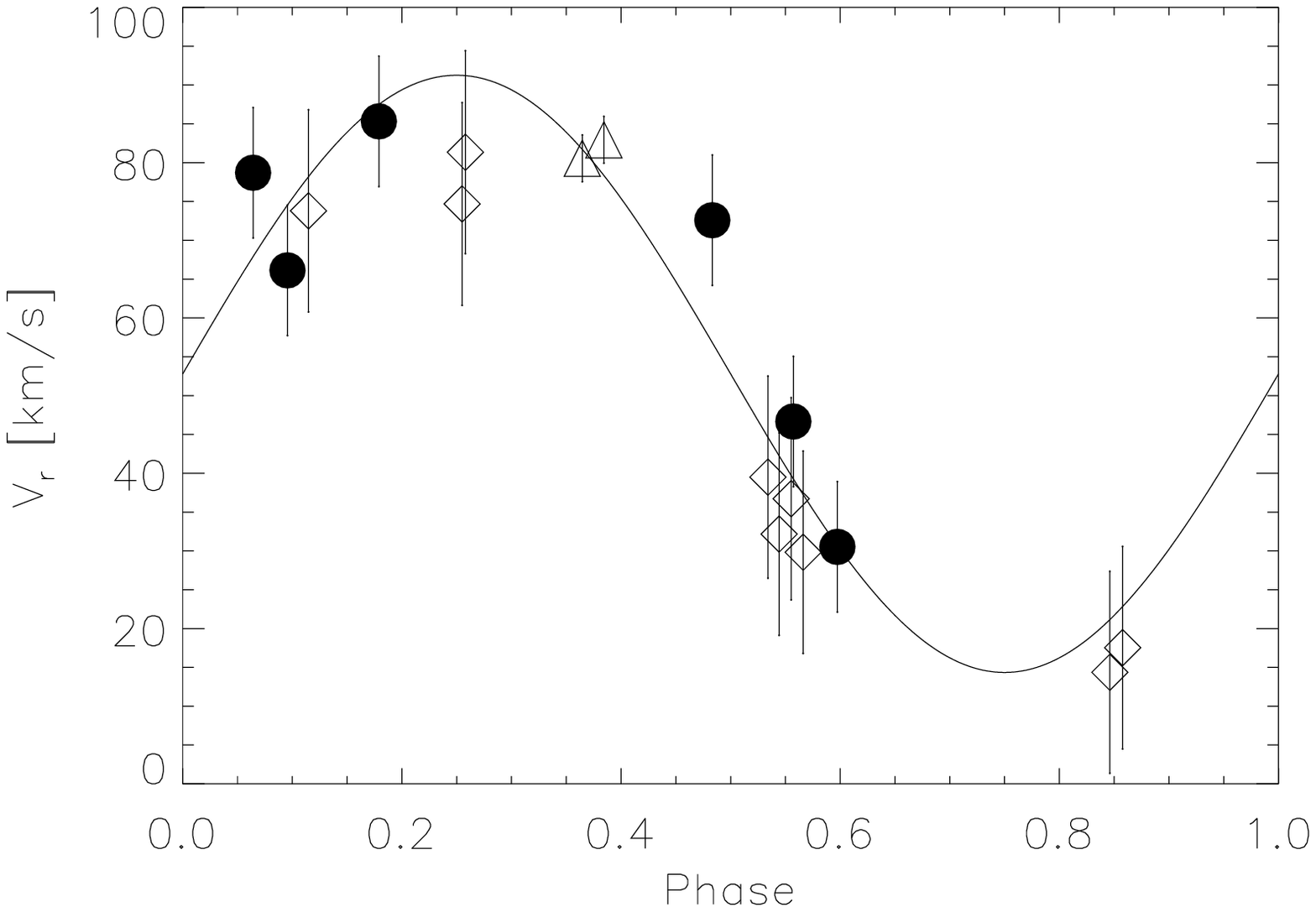}{spec}{Left panel: GMOS-S spectrum of
J0247$-$25. Right panel: Radial velocities of J0247$-$25\,A with a circular
orbit fit. The spectrograph used is indicated as follows: filled circles --
EFOSC2; triangles -- ISIS; diamonds -- GMOS. }

\articlefigure{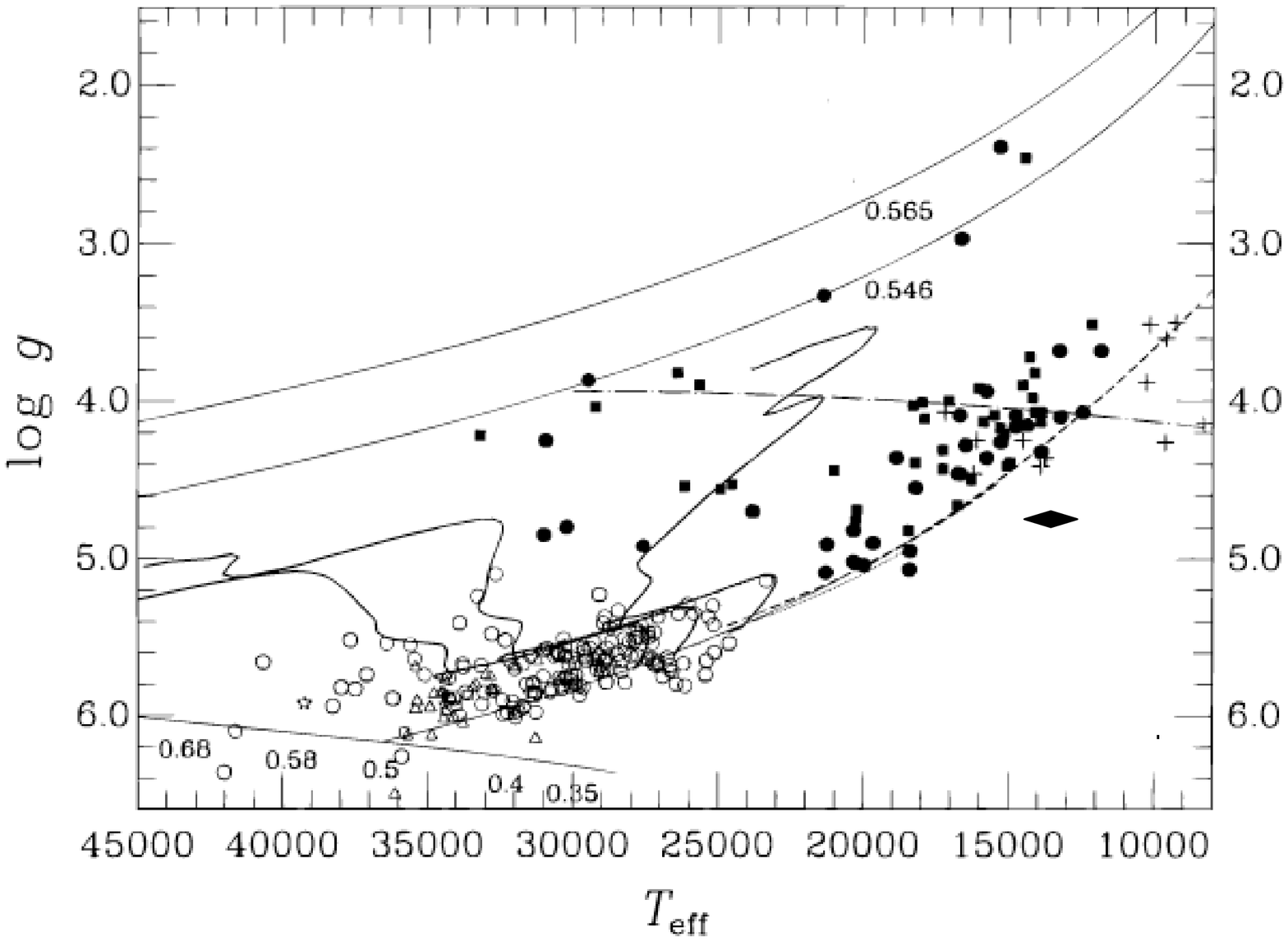}{SafferTlogg}{Location of the J0247$-$25\,B in
the T$_{\rm eff}$ -- $\log g$ plane (solid diamond) compared to 298 faint blue
stars from the survey of \cite{1997ApJ...491..172S}.}

\articlefigure{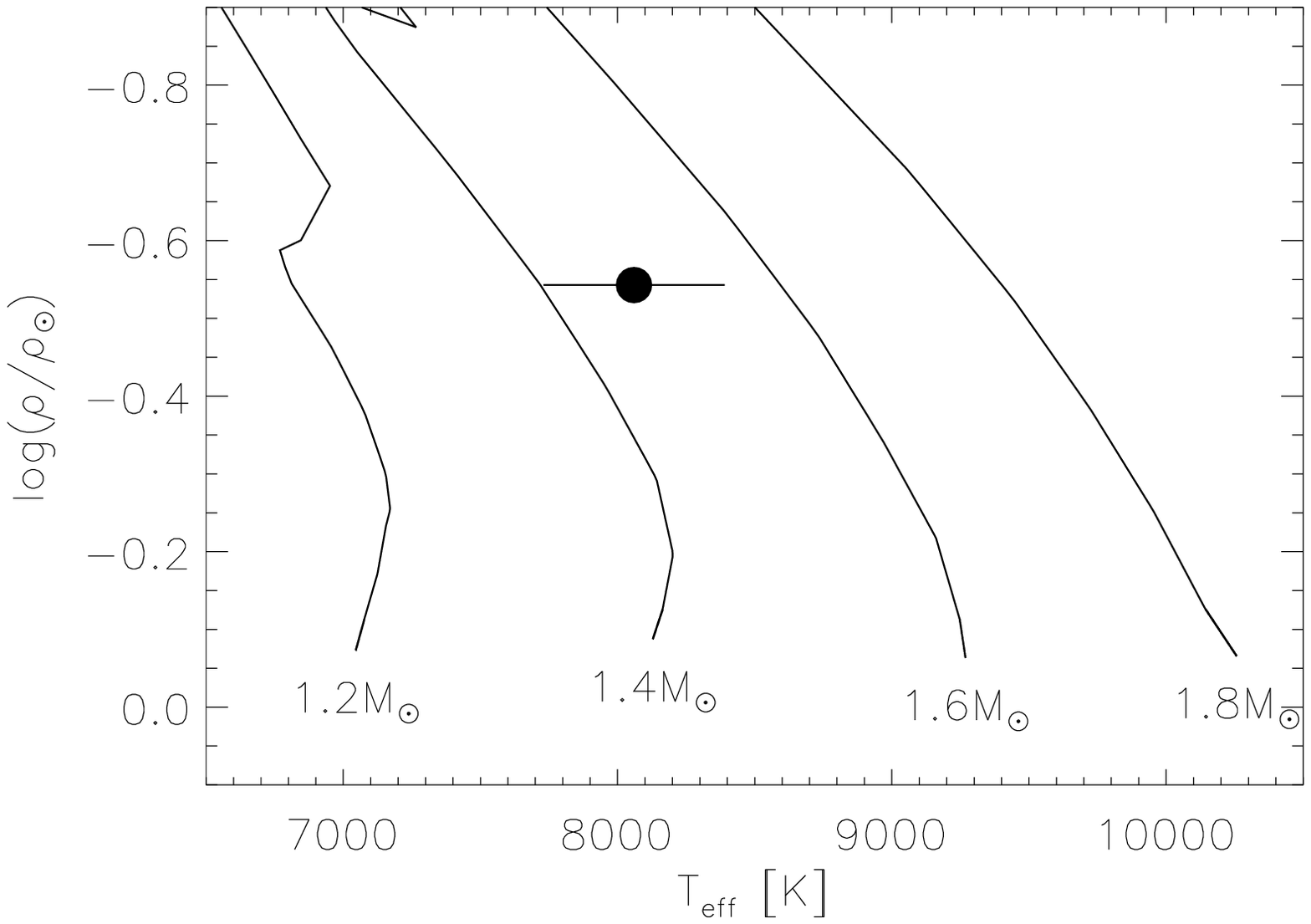}{trho}{Location of the J0247$-$25\,A in the T$_{\rm
eff}$ -- density ($\rho$) plane compared to evolutionary models for normal
stars with masses as noted and [Fe/H]= $-$0.65 from
\cite{2000A&AS..141..371G}. } 

\articlefigure{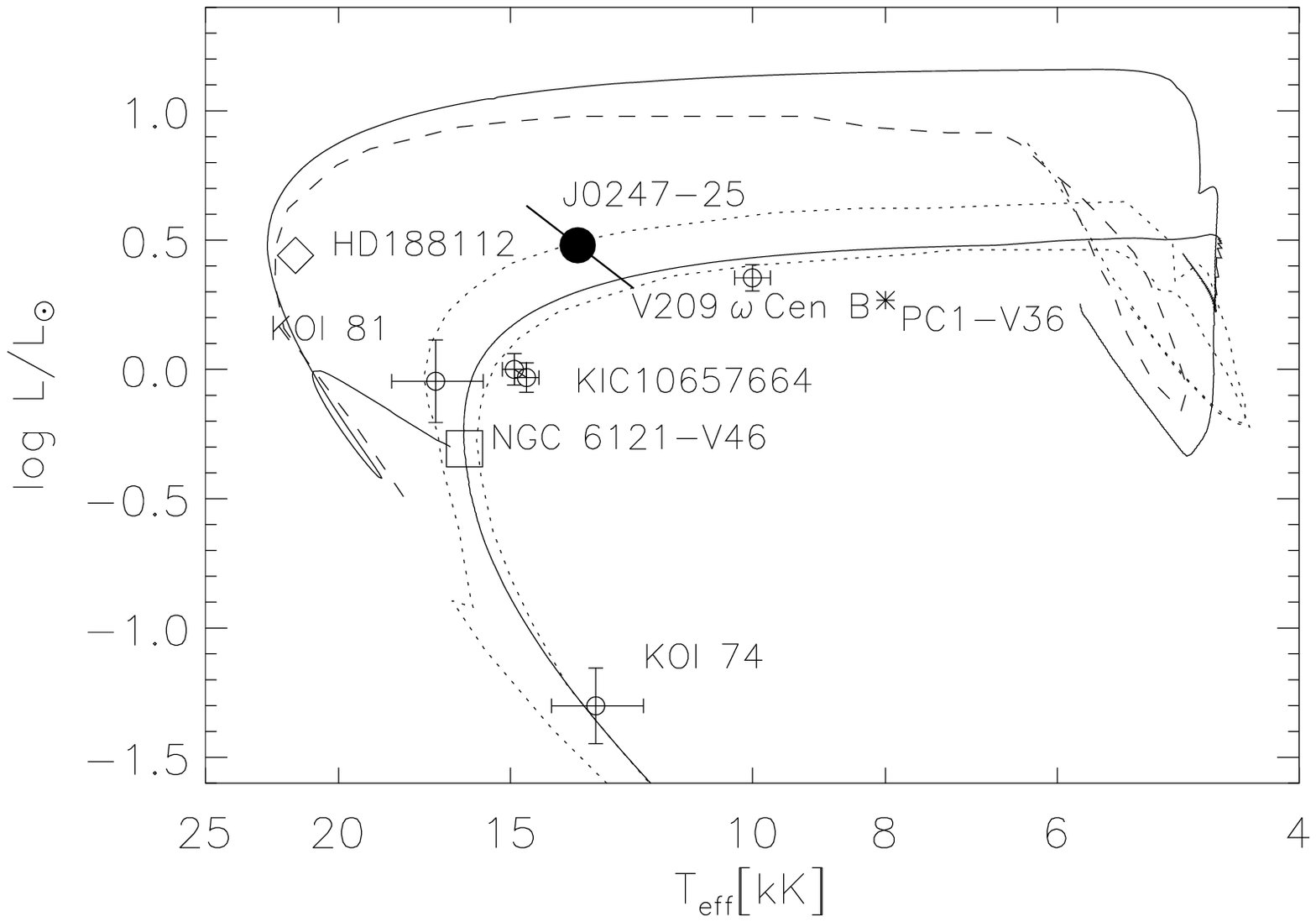}{hrd}{Location of the J0247$-$25\,B and related objects
in the Hertzsprung--Russell diagram. Models for the formation of low mass white
dwarfs (with final masses as noted, bottom-to-top)  are also shown as follows:
\citet{1999A&A...350...89D} -- solid lines (0.195M$_{\odot}$\ and
0.234M$_{\odot}$); \citet{2004ApJ...616.1124N} -- dotted lines
(0.205M$_{\odot}$\ and 0.215M$_{\odot}$); \citet{2010ApJ...715...51V} --
dashed lines (0.21M$_{\odot}$).  }

 The kinematics of J0247$-$25 show that it is a member of the Galactic thick
disk, which suggests that it is likely to be old ($\ga 7$\,Gyr), metal poor
($-1 \la [{\rm Fe/H}] \la -0.3$) and have enhanced $\alpha$-element abundance
([Mg/Fe] $\ga 0.3$). The density of J0247$-$25\,A, $\rho_A =
0.29\pm0.02\,\rho_{\odot}$, can be derived directly from the parameters of the
lightcurve model and the mass function. We compare the values of T$_{\rm
eff,A}$ and  $\rho_A $ to stellar models for the appropriate composition in
Fig.~\ref{trho}. This comparison leads to an estimate of $M_A =
1.5\pm0.1M_{\odot}$ for the mass of J0247$-$25\,A and, via the mass function,
a mass estimate of $M_B= 0.23\pm 0.03M_{\odot}$ for J0247$-$25\,B.

 In Fig.~\ref{hrd} we compare the position of J0247$-$25\,B in the
Hertzsprung-Russell diagram to evolutionary tracks for the formation of low
mass white dwarfs (M$\approx 0.2M_{\odot}$) as a result of drastic mass loss
from low mass red giant stars. The observed properties of J0247$-$25\,B are
well matched by such models during the phase when the star is evolving
bluewards  at almost constant luminosity due to p-p shell-hydrogen burning in
the thin hydrogen envelope. In this scenario J0247$-$25\,B will become a low
mass white dwarf composed almost entirely of helium, so we dub it a pre helium
white dwarf (pre-He-WD).  Also shown in Fig.~\ref{hrd} are other He-WD and
pre-He-WD. The parameters of the related objects are listed in
Table~\ref{lmwd}. 

 A complete description of the discovery and characterisation of J0247$-$25 has
been accepted for publication in MNRAS \citep{2011arXiv1107.4986M}.

\articlefigure{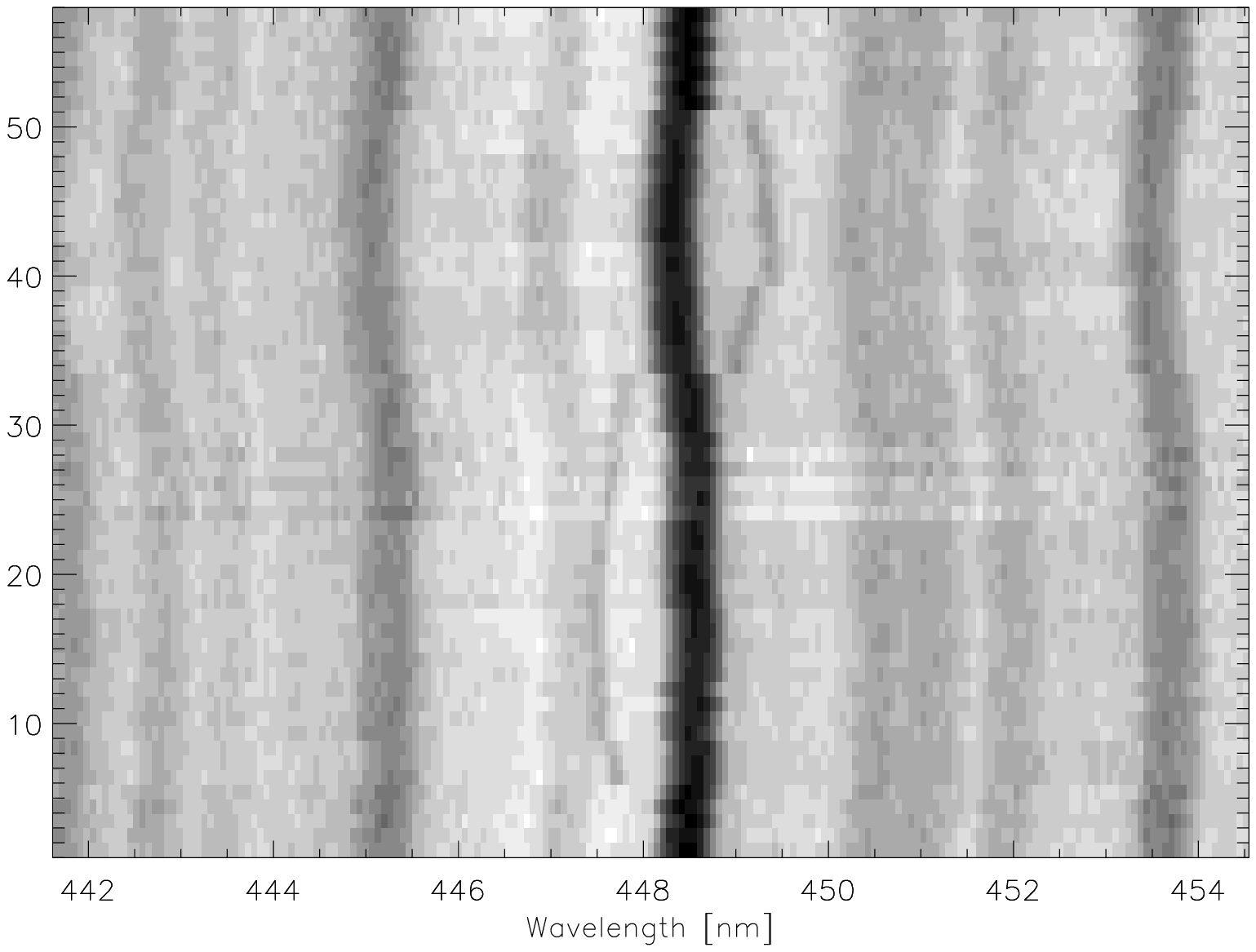}{MgII}{A small section of our UVES spectra of
J0247$-$25 displayed as greyscale images in phase order.}

\articlefigure[angle=270,width=13.4cm]{hgamma.ps}{hgamma}{Spectra of J0247$-$25\,A (upper spectrum)
and J0247$-$25\,B (lower spectrum) around H$\gamma$.} 
\section{UVES spectroscopy of J0247$-$25}

 We obtained high resolution, high signal-to-noise spectroscopy of J0247$-$25
with the UVES echelle spectrograph on the VLT 8.2-m UT2 telescope. Service
mode observations were used to obtain 46 spectra covering the quadrature
phases of the orbit and 12 spectra during the total eclipse, i.e., spectra of
J0247$-$25\,A alone. A small section of these spectra around the Mg\,II
4481\AA\ line is shown in Fig.~\ref{MgII}. The narrow Mg\,II line from
J0247$-$25\,B can be seen moving in anti-phase to the broader and stronger
spectral lines of J0247$-$25\,A.  The mean spectrum obtained during total
eclipse was used as a template to measure the radial velocity of J0247$-$25\,A
by cross correlation. We then subtracted the mean spectrum of J0247$-$25\,A
from the 46 out-of-eclipse spectra after shifting it by the appropriate radial
velocity and scaling it by an estimate of the luminosity ratio at this
wavelength. This process revealed the underlying spectrum of J0247$-$25\,B. We
measured the radial velocity of J0247$-$25\,B using a gaussian fit to the
Mg\,II~4481\AA\ line in these spectra. We then shifted and added these spectra
to produce the mean spectrum of J0247$-$25\,B  shown in Fig.~\ref{hgamma}.
Also visible in this spectrum are a weak He\,I~4471\AA\ line and the broad
H$\gamma$ line.
 
 The radial velocites measured from these UVES spectra combined with the
inclination from the lightcurve model imply masses of $M_A = 2.07 \pm
0.015M_{\odot}$ and $M_B = 0.29 \pm 0.005M_{\odot}$. The coverage of the UVES
spectra is greater than the limited results presented here and several other
spectral lines from J0247$-$25\,B  are visible, so it will be possible to
further improve these mass estimates. Even so, it is clear that the masses of
both J0247$-$25\,A and J0247$-$25\,B are larger than expected based on the
stellar models we have used above. For J0247$-$25\,A the discrepancy between
the mass observed and that expected based on stellar models is similar to that
observed by \citet{2007AJ....133.2457K} for V209~$\omega$~Cen\,A, the
companion to the pre-He-WD V209~$\omega$~Cen\,B in an eclipsing binary member
of the globular cluster $\omega$~Cen.

 We have measured a projected rotational velocity of $V_{\rm rot}\sin i =
95\pm5{\rm km\,s}^{-1} $ for  J0247$-$25\,A from the rotational broadening of
its spectral lines. With our improved mass estimates from the UVES spectroscopy
we find that this is approximately half the rotational velocity expected if
J0247$-$25\,A rotates synchronously with the orbit.

\begin{table}[!ht]
 \caption{Masses and periods for low mass white dwarfs and pre-He-WDs in
binary systems.  } 
 \label{lmwd}
\begin{center}
{\small
 \begin{tabular}{lrrl}
\tableline
\noalign{\smallskip}
  Name &\multicolumn{1}{l}{Period [d]} & \multicolumn{1}{l}{Mass[M$_{\odot}$]} &  Source \\
\noalign{\smallskip}
\tableline
\noalign{\smallskip}
NGC 6121-V46 & 0.087 & $\sim 0.19$ & \citet{2006BaltA..15...61O} \\
HD 188112    & 0.607 & $0.24^{+0.10}_{-0.07} $ & \citet{2003A+A...411L.477H} \\
J0247$-$25\,B & 0.668 & $0.26 \pm 0.03 $&  This paper\\
PC1-V36      & 0.794 &  $0.056 \pm 0.018$ & \citet{2007AJ....133.2457K} \\
V209~$\omega$~Cen\,B & 0.834  & $0.144 \pm 0.008$&\citet{2007AJ....133.2457K}\\
KIC 10657664 & 3.274 &  $0.26 \pm 0.04$ & \citet{2011ApJ...728..139C} \\
             &        &  $0.37 \pm 0.08$ & \multicolumn{1}{c}{"} \\
KOI-75 & 5.189 & $0.22 \pm 0.03$ &  \citet{2010ApJ...715...51V} \\
KOI-81 & 23.89 & $\sim 0.3$      &  \citet{2010ApJ...715...51V} \\
Regulus B    & 40.11  & $> 0.30$  & \citet{2008ApJ...682L.117G} \\
\noalign{\smallskip}
\tableline
\end{tabular}
\newline
}
\end{center}
\end{table}
\section{New eclipsing pre-He-WD}
 We have inspected several thousand lightcurves of stars flagged as eclipsing
binary stars in the WASP archive to look for new examples of eclipsing
pre-He-WD similar to J0247$-$25. The features we looked for in the lightcurve
were: a total eclipse with a depth of about 10\%; sharp ingress/egress to the
total eclipse; a visible secondary eclipse. The lightcurves of 6 stars
satisfying these criteria, including J0247$-$25, are shown in
Fig.~\ref{montage}. The properties of these stars are given in
Table~\ref{new}. The spectral type of the stars has been estimated from the
catalogue photometry available for these stars. Only J1323+43 (EL CVn) has
been previously identified as an eclipsing binary star
\citep{2002MNRAS.331...45K}.

For 4 of these stars we have obtained two spectra with the
TWIN spectrograph on the CAHA 3.5-m telescope, one at each of the quadrature
phases. Observations were obtained with a low resolution grating on the blue arm
covering the wavelength range 3290\,--\,5450\AA. These spectra have been used
to confirm that the spectral types given in Table~\ref{new} are approximately
correct but have not yet been analysed any further. The red arm observations
have a resolution of approximately 1.5\AA\ and cover the H$\alpha$ lines. We
used least squares fitting to determine an empirical line profile for the 
H$\alpha$ line in each star composed of the sum of 3 gaussian functions. This
empirical line profile was then used to measure the radial velocity of the
star at the two quadrature phases observed. We then assumed a mass for the
brighter component of each binary based on its spectral type and used the mass
function to estimate the masses for the fainter pre-He-WD components given in
Table~\ref{new}.  

\section{Discusssion}

 J0247$-$25\,B is an ideal system for testing in detail models for the
formation of low mass helium white dwarfs. It is a bright star, much brighter
than the more distant examples of pre-He-WD found in globular clusters. It
is a double-lined  eclipsing binary star and so it is possible to measure
precise, model-independent masses and radii for both stars in the binary. This
is not possible for most of the other pre-He-WD listed in
Table~\ref{lmwd}. The total eclipses and moderate luminosity ratio of this
binary  make it possible to recover a high quality spectrum of the pre-He-WD
in this binary, as we have shown for our  UVES spectra. This will make it
possible to measure properties of J0247$-$25\,B such as it rotational
velocity, effective temperature  and surface composition. This may make it
possible to test the prediction of some evolutionary models that objects such
as J0247$-$25\,B should be hydrogen deficient. Some low mass white dwarfs are
expected to undergo a number of unstable flashes of CNO hydrogen burning
during their early evolution. The occurance of these flashes depends
critically on the mass of hydrogen that remains on the surface of the star,
which in turn depends on the mass loss history of the star. Understanding
these hydrogen shell flashes is crucial for a better understanding of all low
mass white dwarfs, particularly the low mass white dwarf companions to
millisecond pulsars. It may be possible to put useful constraints on the
hydrogen envelope mass in J0247$-$25\,B by comparing its total mass to the
core mass inferred from its luminosity. The sub-synchronous rotation of
J0247$-$25\,A is rather surprising given that it is expected to have gained
rather a lot of mass and angular momentum from the the red giant progenitor of
J0247$-$25\,B. It may be that this star is currently far from equilibrium. A
detailed reconstruction of the evolutionary history of J0247$-$25 will lead to
a much better understanding of how stars react to mass accretion. This will
obviously be interesting for improving our understanding of binary star
evolution, but may have wider implications, e.g., episodic accretion may be
the process that dominates the observed properties of pre main-sequence stars
\citep{2009ApJ...702L..27B}. The kinematics of J0247$-$25 also imply useful
constraints on the composition and age of this binary star.

 The discovery of several other eclipsing pre-He-WD opens up the possibility
of exploring how the formation of these objects varies with parameters such as
the initial masses and orbital periods of the binary. It also makes the tests
of the evolution models for these objects much stronger because fine tuning of
parameters or extraordinary evolutionary scenarios that might be invoked to
explain the formation of a single object cannot be justified when several
similar examples exist. It may also be possible to put useful constraints on
the space density of such objects since the WASP survey seems to be very
effective at detecting these short period eclipsing binaries.

\articlefigure{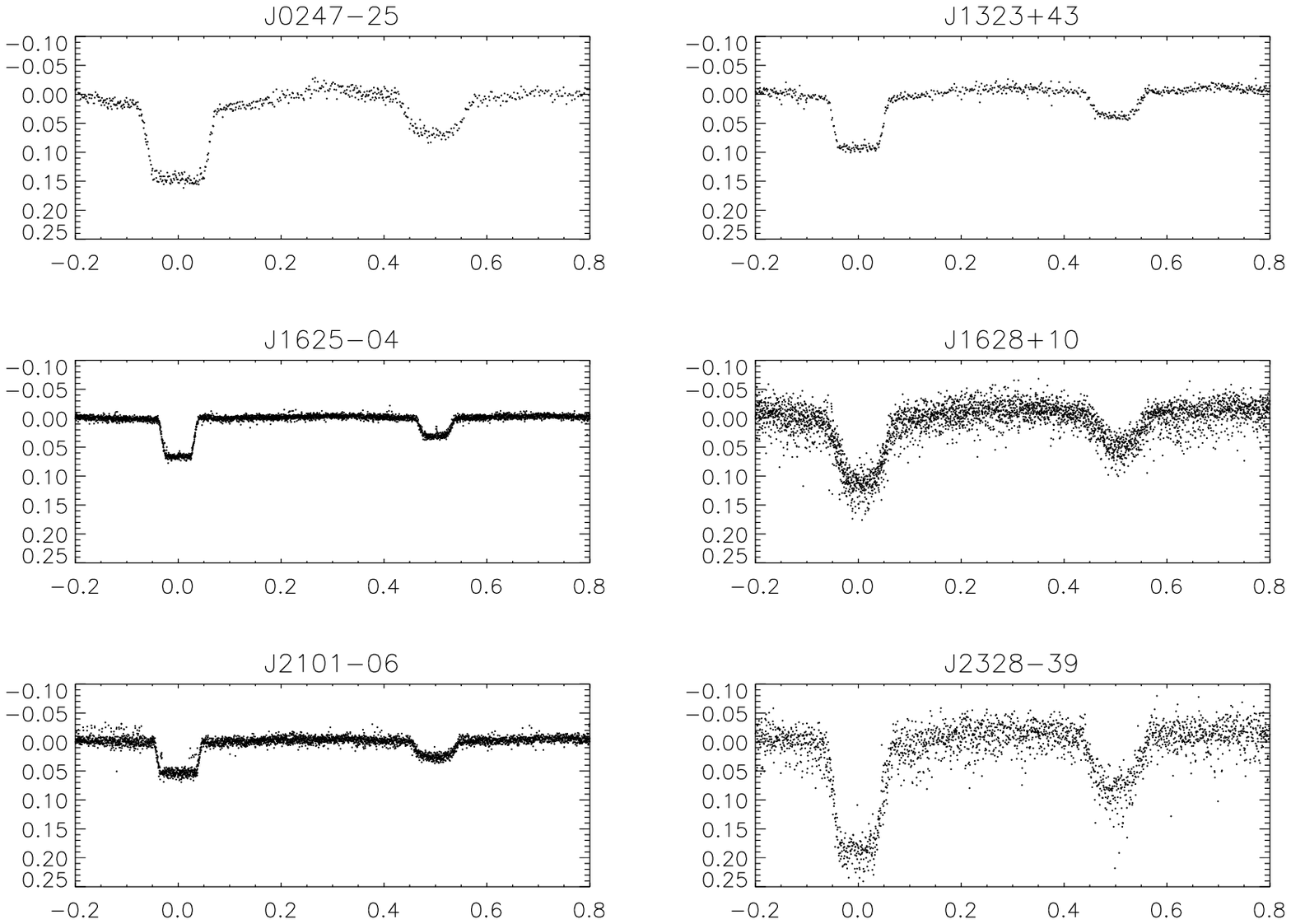}{montage}{WASP lightcurves of eclipsing pre-HE-WD.}

\begin{table}[!ht]
\caption{Newly identified eclipsing pre-He-WD}
\label{new}
\smallskip
\begin{center}
{\small
\begin{tabular}{lrrrr}
\tableline
\noalign{\smallskip}
Star & Spectral & \multicolumn{1}{c}{V}  & Period  & 
\multicolumn{1}{c}{M$_2$}  \\
&\multicolumn{1}{l}{Type}& [mag] & \multicolumn{1}{c}{[d]} & [M$_{\odot}$] \\
\noalign{\smallskip}
\tableline
\noalign{\smallskip}
J0247$-$25 & A6 & 11.9 & 0.668 & 0.29       \\
J1323+43   & A1 &  9.4 & 0.795 & $\sim$ 0.2 \\
J1625$-$04 & A7 & 10.4 & 1.526 & $\sim$ 0.15 \\
J1628+10   & F6 & 12.9 & 0.720 & $\sim$ 0.05 \\
J2101$-$06   & A2 & 11.5 & 1.290 & $\sim$ 0.2 \\
J2328$-$39   & A6 & 13.3 & 0.769 & -- \\
\noalign{\smallskip}
\tableline
\end{tabular}
}
\end{center}
\end{table}

\acknowledgements 
 Funding for WASP comes from consortium universities and from the UK's
Science and Technology Facilities Council (STFC).
 The research leading to these results has received funding from the European
Research Council under the European Community's Seventh Framework Programme
(FP7/2007--2013)/ERC grant agreement n$^\circ$227224 (PROSPERITY), as well as
from the Research Council of K.U.Leuven grant agreement GOA/2008/04.

\bibliography{maxted}

\begin{thebibliography}{}
\expandafter\ifx\csname natexlab\endcsname\relax\def\natexlab#1{#1}\fi
\expandafter\ifx\csname url\endcsname\relax
  \def\url#1{\texttt{#1}}\fi
\expandafter\ifx\csname urlprefix\endcsname\relax\def\urlprefix{URL }\fi
\providecommand{\eprint}[2][]{\url{#2}}

\bibitem[{{Baraffe} et~al.(2009){Baraffe}, {Chabrier}, \&
  {Gallardo}}]{2009ApJ...702L..27B}
{Baraffe}, I., {Chabrier}, G., \& {Gallardo}, J. 2009, \apjl, 702, L27.
  \eprint{0907.3886}

\bibitem[{{Carter} et~al.(2011){Carter}, {Rappaport}, \&
  {Fabrycky}}]{2011ApJ...728..139C}
{Carter}, J.~A., {Rappaport}, S., \& {Fabrycky}, D. 2011, \apj, 728, 139.
  \eprint{1009.3271}

\bibitem[{{Driebe} et~al.(1999){Driebe}, {Bl{\"o}cker}, {Sch{\"o}nberner}, \&
  {Herwig}}]{1999A&A...350...89D}
{Driebe}, T., {Bl{\"o}cker}, T., {Sch{\"o}nberner}, D., \& {Herwig}, F. 1999,
  \aap, 350, 89. \eprint{arXiv:astro-ph/9908156}

\bibitem[{{Etzel}(1981)}]{1981psbs.conf..111E}
{Etzel}, P.~B. 1981, in Photometric and Spectroscopic Binary Systems, edited by
  {E.~B.~Carling \& Z.~Kopal}, 111

\bibitem[{{Gies} et~al.(2008){Gies}, {Dieterich}, {Richardson}, {Riedel},
  {Team}, {McAlister}, {Bagnuolo}, {Grundstrom}, {{\v S}tefl}, {Rivinius}, \&
  {Baade}}]{2008ApJ...682L.117G}
{Gies}, D.~R., {Dieterich}, S., {Richardson}, N.~D., {Riedel}, A.~R., {Team},
  B.~L., {McAlister}, H.~A., {Bagnuolo}, W.~G., Jr., {Grundstrom}, E.~D., {{\v
  S}tefl}, S., {Rivinius}, T., \& {Baade}, D. 2008, \apjl, 682, L117.
  \eprint{0806.3473}

\bibitem[{{Girardi} et~al.(2000){Girardi}, {Bressan}, {Bertelli}, \&
  {Chiosi}}]{2000A&AS..141..371G}
{Girardi}, L., {Bressan}, A., {Bertelli}, G., \& {Chiosi}, C. 2000, \aaps, 141,
  371. \eprint{arXiv:astro-ph/9910164}

\bibitem[{{Heber} et~al.(2003){Heber}, {Edelmann}, {Lisker}, \&
  {Napiwotzki}}]{2003A+A...411L.477H}
{Heber}, U., {Edelmann}, H., {Lisker}, T., \& {Napiwotzki}, R. 2003, \aap, 411,
  L477

\bibitem[{{Kaluzny} et~al.(2007){Kaluzny}, {Rucinski}, {Thompson}, {Pych}, \&
  {Krzeminski}}]{2007AJ....133.2457K}
{Kaluzny}, J., {Rucinski}, S.~M., {Thompson}, I.~B., {Pych}, W., \&
  {Krzeminski}, W. 2007, \aj, 133, 2457. \eprint{0704.3507}

\bibitem[{{Koen} \& {Eyer}(2002)}]{2002MNRAS.331...45K}
{Koen}, C., \& {Eyer}, L. 2002, \mnras, 331, 45.
  \eprint{arXiv:astro-ph/0112194}

\bibitem[{{Maxted} et~al.(2011){Maxted}, {Anderson}, {Burleigh},
  {Collier-Cameron}, {Heber}, {Gaensicke}, {Geier}, {Kupfer}, {Marsh},
  {Nelemans}, {O'Toole}, {Ostensen}, {Smalley}, \&
  {West}}]{2011arXiv1107.4986M}
{Maxted}, P.~F.~L., {Anderson}, D.~R., {Burleigh}, M.~R., {Collier-Cameron},
  A., {Heber}, U., {Gaensicke}, B.~T., {Geier}, S., {Kupfer}, T., {Marsh},
  T.~R., {Nelemans}, G., {O'Toole}, S.~J., {Ostensen}, R.~H., {Smalley}, B., \&
  {West}, R.~G. 2011, Accepted for publication in MNRAS. \eprint{1107.4986}

\bibitem[{{Nelson} et~al.(2004){Nelson}, {Dubeau}, \&
  {MacCannell}}]{2004ApJ...616.1124N}
{Nelson}, L.~A., {Dubeau}, E., \& {MacCannell}, K.~A. 2004, \apj, 616, 1124

\bibitem[{{O'Toole} et~al.(2006){O'Toole}, {Napiwotzki}, {Heber}, {Drechsel},
  {Frandsen}, {Grundahl}, \& {Bruntt}}]{2006BaltA..15...61O}
{O'Toole}, S.~J., {Napiwotzki}, R., {Heber}, U., {Drechsel}, H., {Frandsen},
  S., {Grundahl}, F., \& {Bruntt}, H. 2006, Baltic Astronomy, 15, 61.
  \eprint{arXiv:astro-ph/0605441}

\bibitem[{{Pollacco} et~al.(2006)}]{2006PASP..118.1407P}
{Pollacco}, D.~L., et~al. 2006, \pasp, 118, 1407.
  \eprint{arXiv:astro-ph/0608454}

\bibitem[{{Popper} \& {Etzel}(1981)}]{1981AJ.....86..102P}
{Popper}, D.~M., \& {Etzel}, P.~B. 1981, \aj, 86, 102

\bibitem[{{Saffer} et~al.(1997){Saffer}, {Keenan}, {Hambly}, {Dufton}, \&
  {Liebert}}]{1997ApJ...491..172S}
{Saffer}, R.~A., {Keenan}, F.~P., {Hambly}, N.~C., {Dufton}, P.~L., \&
  {Liebert}, J. 1997, \apj, 491, 172

\bibitem[{{van Kerkwijk} et~al.(2010){van Kerkwijk}, {Rappaport}, {Breton},
  {Justham}, {Podsiadlowski}, \& {Han}}]{2010ApJ...715...51V}
{van Kerkwijk}, M.~H., {Rappaport}, S.~A., {Breton}, R.~P., {Justham}, S.,
  {Podsiadlowski}, P., \& {Han}, Z. 2010, \apj, 715, 51. \eprint{1001.4539}

\end{thebibliography}

\end{document}